# Optimal Relay Selection and Resource Allocation for Two Way DF-AF Cognitive Radio Networks

Ahmad Alsharoa, Faouzi Bader**, and Mohamed-Slim Alouini,

Computer, Electrical, and Mathematical Science of Engineering (CEMSE) Division,

King Abdullah University of Science and Technology (KAUST),

Thuwal, Makkah Province, Kingdom of Saudi Arabia,

** Centre Tecnologic de Telecommunication de Catalunya (CTTC),

Parc Mediterrani de la Tecnologia, Av. Carl Friedrich Gauss 7, 08860, Castelldefels-Barcelona, Spain,

E-mails: {ahmad.sharoa, slim.alouini}@kaust.edu.sa; {faouzi.bader}@cttc.es.

**Abstract**

In this letter, the problem of optimal resource power allocation and relay selection for two way relaying cognitive radio networks using half duplex Decode and Forward (DF) and Amplify and Forward (AF) systems are investigated. The primary and secondary networks are assumed to access the spectrum at the same time, so that the interference introduced to the primary network caused by the secondary network should be below a certain interference threshold. In addition, a selection strategy between the AF and DF schemes is applied depending on the achieved secondary sum rate without affecting the quality of service of the primary network. A suboptimal approach based on a genetic algorithm is also presented to solve our problem. Selected simulation results show that the proposed suboptimal algorithm offers a performance close to the performance of the optimal solution with a considerable complexity saving.

## I. INTRODUCTION

Cognitive Radio (CR) has recently attracted enormous attention in wireless communication networks [1]. It is considered as a promising solution towards a more efficient usage of the radio spectrum. The idea of CR spectrum sharing is to allow unlicensed users known also as Secondary Users (SUs) to utilize the spectrum band allocated by licensed users known also as Primary Users (PUs) at the same time. In order to protect the PUs, the interference due to the SUs should be kept under a certain interference temperature limit.

On another front, there has been recently a great deal of interest in two way relaying networks [2], [3]. The transmission process in this relaying technique takes place in two time slots. In the first slot, the source and the destination transmit their signals simultaneously to the relay. Subsequentaly, in the second slot, the relay broadcasts its signal to the terminals. Two widely relay protocols are used in practice, the namely Amplify and Forward (AF) protocol, which amplifies the received signal first, then broadcast it to the destination, and the Decode and Forward (DF) protocol, which decodes the received signal to remove the noise before transmitting a clean copy of the original signal to the destination. For instance, the work presented by Chen *et. al* in [2] deals with multi access two way relaying network case, while in [3] the authors show analytically and via simulation that two way relaying outperforms one way relaying in terms of energy efficiency. Furthermore, the relay selection and power allocation problems for AF protocol in cooperative one way and two way relaying CR have been investigated in [4] and [5], respectively. The best relay selection in two way relaying depends on two factors, end to end channel conditions, and the presence of the Primary Network (PN) according to the interference constraints . On the other hand, prior work in the literature has studied adaptive relaying which allows the switching between AF and DF protocols depending on the Signal-to-Noise-Ratio (SNR) [6]. However, to the best knowledge of the authors, the relay selection problem in two way relaying CR networks using DF protocol has not been discussed so far as it is the case for the AF protocol.

In this letter, a best relay selection scheme for two way relaying CR with half duplex case and channel reciprocity is considered. In the AF protocol, the relay broadcasts the amplified copy of the received signal to the terminals, i.e., the noise gets amplified too. On the other hand, in the DF protocol, the relay regenerates clean signals from the received signal and transmits the re-encoded message to the terminals. More specifically, the main contributions for our new proposed scheme can be summarized as follows:

- Formulate a new relay selection scheme in two way relaying CR system which selects between the DF and AF protocols depending on the higher Sum Rate (SR) achieved by the

Secondary Network (SN) without affecting the Quality of Service (QoS) of the PN. For that reason, additional interference constraints are considered in the optimization problem for both time slots (it is assumed that $I_{th}$ is the same in each time slot).

- Derivation of the optimal transmits power and relay power that maximize the cognitive SR of the system.
- Using dual decomposition and subgradient methods for both AF and DF techniques in order to solve the SR maximization problem and select the best relay with the best technique.
- Design a practical low complexity suboptimal approach based on Genetic Algorithm (GA) to solve the formulated optimization problem [7], and compare it with the optimal and Exhaustive Search (ES) solutions.

Generally, in one way relaying, it is assumed that at high SNR the relay can decode perfectly, so it achieves higher SR using the DF protocol. On the contrary, for low SNR the higher SR can be achieved using the AF protocol. However, the results provided in Section V show that in two way relaying at high SNR the DF protocol becomes as a bottleneck in the first phase, so higher SR can be achieved using the AF protocol. On the other hand, for low SNR, the relay with the DF protocol achieves higher SR.

The rest of this letter is organized as follows. Section II gives the system model. The problem formulation and the optimal algorithm are described in Section III. The suboptimal scheme is presented in Section IV. Simulations and numerical results are demonstrated in Section V. Finally, the letter is concluded in Section VI.

## II. SYSTEM MODEL

In this section, the best relay selection problem for CR two way relaying is investigated. The SN is constituted of a cognitive Mobile User (MU), a Cognitive Base station (CB), and $M$ Relay Stations (RSs) as illustrated in Fig.1. It is assumed that there is no direct link between the cognitive terminals and the single relay principle is applied to select the best relay. During the first time slot known also as the Multiple Access Channel (MAC) phase, the CB transmits its signal to RSs with power denoted $P_{CB}$. Concurrently the secondary MU transmits its signal to RSs with power denoted $P_S$. This causes two interferences to the PU. In the second time slot known also as the Broadcast Channel phase (BC), the selected RS broadcast its signal. This phase also causes interference to the PU from the active RS.

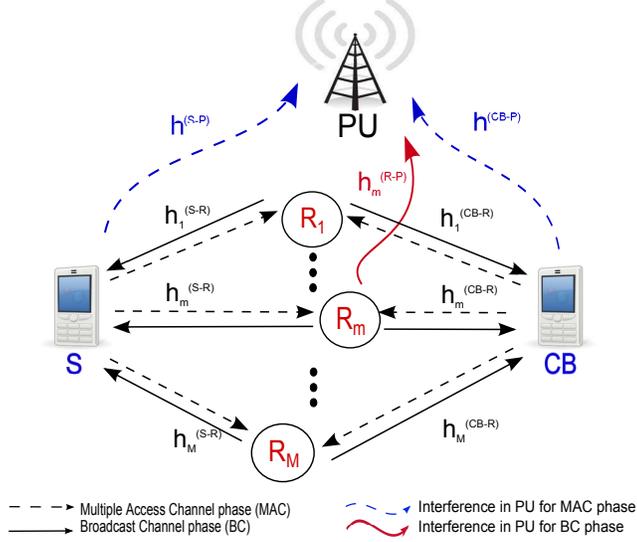

Fig. 1. System model of the cooperative two way relaying cognitive radio system.

We assume that all the channel gains are perfectly known at the communication nodes. All channel gains for the network can be adopted by assuming channel reciprocity and classical channel estimation approaches [8]. The interference between the PN and SN is studied in Section III. Also, we assume that the PN and SN access the spectrum at the same time. Without loss of generality, all the noise variances are assumed to be equal to $\sigma_n^2$. Finally, selection strategy between the AF and the DF protocols is applied in order to achieve the maximum SR of the SN without affecting the QoS of the PU measured by $I_{th}$.

## III. PROBLEM FORMULATION

Let us define $R_{DF}$ and $R_{AF}$ as the achievable secondary SR for the DF protocol, and the achievable secondary SR for the AF protocol, respectively.

The Optimization Problem1 (OP1) for a single relay selection can be formulated as[1]

$$\text{(OP1):} \quad m^* = \underset{m \in \{1:M\}}{\operatorname{argmax}} \; \underset{R}{\max} \; \rho R_{DF} + (1-\rho) R_{AF}, \tag{1}$$

$$s.t \quad 0 \leq P_S \leq \bar{P}_S, \tag{2}$$

$$0 \leq P_{CB} \leq \bar{P}_{CB}, \tag{3}$$

---

[1]For simplicity and uniformity we use the mathematical notations depicted in Table I.

$$0 \leq P_{R,m} \leq \bar{P}_R, \quad \forall m = 1, ..., M, \qquad (4)$$

- interference constraint in first time slot

$$\left[ f_3 P_S + f_4 P_{CB} \right] \leq I_{th}, \qquad (5)$$

- interference constraint in second time slot

$$f_5 P_{R,m} \leq I_{th}, \quad \forall m = 1, ..., M, \qquad (6)$$

where $\bar{P}_S, \bar{P}_{CB}$, and $\bar{P}_R$, are the peak transmit power of the secondary MU, CB, and $m$-th RS, respectively. In (1), $\rho$ is a constant equal to either zero for the AF protocol or one for the DF protocol, and the channels coefficients are given in Table I. Let $x_1$ and $x_2$ are the symbols transmitted by the MU and CB respectively. It is assumed that $\mathbb{E}(|x_1|^2) = \mathbb{E}(|x_2|^2) = 1^2$. In the first time slot, the received signal at the $m$-th relay is given by

$$r_m = \sqrt{P_S} h_m^{(S-R)} x_1 + \sqrt{P_{CB}} h_m^{(CB-R)} x_2 + z_m, \qquad (7)$$

where $z_m$ is the additive Gaussian noise at the $m$-th relay.

TABLE I: Symbol Notations

| Symbol | Notation | Complex Channel Gain between |
| --- | --- | --- |
| $f_1$ | $|h_m^{(CB-R)}|^2$ | CB and RS $m$ |
| $f_2$ | $|h_m^{(S-R)}|^2$ | MU and RS $m$ |
| $f_3$ | $|h^{(S-P)}|^2$ | MU and PU |
| $f_4$ | $|h^{(CB-P)}|^2$ | CB and PU |
| $f_5$ | $|h_m^{(R-P)}|^2$ | RS $m$ and PU |

In order to simplify the formulated OP1, we solve it time slot per time slot. During the BC phase, the power allocation at the $m$-th relay depends essentially on two constraints: the peak power constraint (4) and the interference constraint (6). For this reason, the optimal relay power can be expressed as

$$P_{R,m}^* = \min\left(\bar{P}_R, \frac{I_{th}}{f_5}\right), \quad \forall m = 1, ..., M. \qquad (8)$$

---

[2]$\mathbb{E}(\cdot)$ denotes the expectation operator.

The optimization problem during the MAC phase is therefore given by

$$\text{(OP2):} \quad m^* = \underset{m \in \{1:M\}}{\operatorname{argmax}} \; \underset{R}{\max} \; \rho R_{DF} + (1-\rho) R_{AF}, \tag{9}$$

$$\text{s.t} \quad (2), (3), (5) \tag{10}$$

We can decompose the OP2 outlined above into parallel subproblems using single relay principle, i.e., each independently solvable for a different relay, then we select the relay that achieves maximum SR.

The dual subproblem associated with OP2 can be written as [9]

$$\underset{\boldsymbol{\lambda} \geq 0}{\min} \; g(\boldsymbol{\lambda}), \tag{11}$$

where $\boldsymbol{\lambda}$ is a lagrangian vector contains the Lagrangian multipliers in the system. The dual function $g(\boldsymbol{\lambda})$ is defined as follows

$$g(\boldsymbol{\lambda}) = \underset{P_S \geq 0, P_{CB} \geq 0}{\max} \; \mathcal{L}(\boldsymbol{\lambda}, P_S, P_{CB}). \tag{12}$$

*A. Amplify and Forward Protocol*

In this protocol, The relay amplifies the received signal by $w_m$, then the received signal at the terminals can be expressed as

$$\begin{aligned} r_{m,S} &= w_m h_m^{(S-R)} r_m + z_{CB}, \\ r_{m,CB} &= w_m h_m^{(CB-R)} r_m + z_S, \end{aligned} \tag{13}$$

where $z_{CB}$ and $z_S$ are the additive Gaussian noise at the terminals. By using the perfect knowledge of the channel gains and channel reciprocity, the terminals can remove the self interference by eliminating their own signals. Thus, the SNR at MU and CB are given by

$$\gamma_{m,S} = \frac{P_{CB}|w_m|^2 f_2 f_1}{\sigma_n^2(|w_m|^2 f_2 + 1)}, \gamma_{m,CB} = \frac{P_S|w_m|^2 f_2 f_1}{\sigma_n^2(|w_m|^2 f_1 + 1)}, \tag{14}$$

respectively. The relay power of the *m*-th relay node can be expressed as

$$P_{R,m} = \mathbb{E}(|w_m r_m|^2) = (P_S f_2 + P_{CB} f_1 + \sigma_n^2)|w_m|^2. \tag{15}$$

By substituting the value of $|w_m|^2$ from (15) into (14), the SNRs become

$$\begin{aligned} \gamma_{m,S} &= \frac{P_{CB} P_{R,m}^* f_2 f_1}{\sigma_n^2(P_{R,m}^* f_1 + P_S f_2 + P_{CB} f_1 + \sigma_n^2)}, \\ \gamma_{m,CB} &= \frac{P_S P_{R,m}^* f_2 f_1}{\sigma_n^2(P_{R,m}^* f_2 + P_S f_2 + P_{CB} f_1 + \sigma_n^2)}. \end{aligned} \tag{16}$$

The achieved SR for AF protocol of two way relaying can be written as

$$R_{AF} = \frac{1}{2}\log_2(1+\gamma_{m,S}) + \frac{1}{2}\log_2(1+\gamma_{m,CB}). \tag{17}$$

Due to the non-convexity of the formula in AF protocol, a convex approximation when the system operates at high SNR region is presented [5]:

$$R_{AF} \approx \frac{1}{2}\log_2(\gamma_{m,S}) + \frac{1}{2}\log_2(\gamma_{m,CB}). \tag{18}$$

When $\rho = 0$ and due the fact that the logarithmic function is a monotonically increasing function of its arguments, the OP2 is equivalent to the following

$$\text{(OP2):} \quad m^* = \underset{m \in \{1:M\}}{\operatorname{argmax}} \quad \min \frac{1}{\gamma_{m,CB} \cdot \gamma_{m,CB}}, \tag{19}$$

$$\text{s.t} \quad (2), (3), (5) \tag{20}$$

Thus, the Lagrangian of AF protocol can be written as

$$\mathcal{L}_{AF} = -\frac{1}{\gamma_{m,CB} \cdot \gamma_{m,CB}} - \lambda_S(P_S - \bar{P}_S) - \lambda_{CB}(P_{CB} - \bar{P}_{CB}) - \lambda_1(f_3 P_S + f_4 P_{CB} - I_{th}), \tag{21}$$

where $\lambda_S, \lambda_{CB}$, and $\lambda_1$ represent the Lagrangian multipliers related to the peak power at the source, peak power at the destination, and interference constraint in the first time slot, respectively. By applying the Karush-Kuhn-Tucker (KKT) optimality conditions [9], we obtain

$$\frac{\partial \mathcal{L}_{AF}}{P_S} = 0 \quad \text{and} \quad \frac{\partial \mathcal{L}_{AF}}{P_{CB}} = 0. \tag{22}$$

Direct calculation yields

$$P_S^* = \sqrt{\left(\frac{\sigma_n^4 A}{\sigma_n^4 f_2^2 + (\lambda_S + \lambda_1 f_3) P_{CB} P_{R,m}^{*2} f_1^2 f_2^2}\right)^+} \tag{23}$$

$$P_{CB}^* = \sqrt{\left(\frac{\sigma_n^4 B}{\sigma_n^4 f_1^2 + (\lambda_{CB} + \lambda_1 f_4) P_S P_{R,m}^{*2} f_1^2 f_2^2}\right)^+} \tag{24}$$

where $A = P_{R,m}^{*2} f_2 f_1 + P_{R,m}^*(f_2 \sigma_n^2 + f_1 \sigma_n^2) + P_{CB}^2 f_1^2 + P_{CB}(2f_1\sigma_n^2) + P_{R,m}^* P_{CB}(f_2 f_1 + f_1^2) + \sigma_n^4$, $B = P_{R,m}^{*2} f_2 f_1 + P_{R,m}^*(f_2 \sigma_n^2 + f_1 \sigma_n^2) + P_S^2 f_2^2 + P_S(2f_2\sigma_n^2) + P_{R,m}^* P_S(f_2 f_1 + f_2^2) + \sigma_n^4$. and $(x)^+$ denotes a maximum between $x$ and zero.

## B. Decode and Forward Protocol

Prior works in the literature have studied the sum rate for two way relaying with DF protocol [10]–[12]. The max SR of the DF protocol can be expressed as

$$R_{DF} = \frac{1}{2}\min\Big[\min\{R_1, R_3\} + \min\{R_2, R_4\}, R_5\Big], \qquad (25)$$

where $R_1 = \log_2\Big(1 + \frac{P_S f_2}{\sigma_n^2}\Big)$, $R_2 = \log_2\Big(1 + \frac{P_{CB} f_1}{\sigma_n^2}\Big)$, denote the rate from the source and the destination to the relay in the first time slot, respectively, $R_3 = \log_2\Big(1 + \frac{P_R f_1}{\sigma_n^2}\Big)$, $R_4 = \log_2\Big(1 + \frac{P_R f_2}{\sigma_n^2}\Big)$, denote the rate from the relay to the source and to the destination in the second time slot, respectively, and $R_5 = \log_2\Big(1 + \frac{P_{CB} f_1 + P_S f_2}{\sigma_n^2}\Big)$ denotes the max SR can be achieved in both time slots.

It is assumed that the relay node decodes the high SNR signal (Down-Link (DL) signal) first, then decodes the other signal (Up-Link (UL) signal) after subtracting the decoded signal. For this reason additional Lagrangian multipliers are considered for UL and DL. When $\rho = 1$, the Lagrangian of OP2 can be written as

$$\mathcal{L}_{\text{DF}} = (1 - \lambda_u - 1 + \lambda_d)\tfrac{1}{2}\log_2(1 + \tfrac{P_S f_2}{\sigma_n^2})(1 - \lambda_d)\tfrac{1}{2}\log_2(1 + \tfrac{P_{CB} f_1 + P_S f_2}{\sigma_n^2}) - \lambda_S(P_S - \bar{P}_S) - \\ \lambda_{CB}(P_{CB} - \bar{P}_{CB}) - \lambda_1(f_3 P_S + f_4 P_{CB} - I_{th}). \qquad (26)$$

where $\lambda_u$ and $\lambda_d$ are the dual variables associated with the UL and DL rate constraints, respectively. Letting $\alpha = 2.\ln 2$ and applying the KKT optimality conditions, we obtain

$$\frac{\partial \mathcal{L}_{\text{DF}}}{P_S} = 0 \quad \text{and} \quad \frac{\partial \mathcal{L}_{\text{DF}}}{P_{CB}} = 0. \qquad (27)$$

Direct calculation yields

$$P_{CB}^* = \left(\frac{(1 - \lambda_d)}{\alpha(\lambda_1 f_4 + \lambda_{CB})} - \frac{P_S f_2 + \sigma_n^2}{f_1}\right)^+ \qquad (28)$$

$$\kappa_1 P_S^{*2} + \kappa_2 P_S^* + \kappa_3 = 0, \qquad (29)$$

where $\kappa_1 = (\lambda_1 f_3 + \lambda_S) f_2^2$, $\kappa_2 = f_2(2\sigma_n^2 + P_{CB} f_1)(\lambda_1 f_3 + \lambda_S) - \frac{(1-\lambda_u)f_2}{\alpha}$, and $\kappa_3 = (1 - \lambda_d) P_{CB} f_1 \frac{f_2}{\alpha} + (\sigma_n^2(\lambda_1 f_3 + \lambda_S) - (1 - \lambda_u)\frac{f_2}{\alpha})(\sigma_n^2 + P_{CB} f_1)$. By substituting (28) into (29) and after simplification, we obtain the optimal source power as the following

$$P_S^* = \left(\frac{(\lambda_d - \lambda_u)(\frac{f_2 f_1 (1-\lambda_d)}{\alpha^2 (\lambda_1 f_4 + \lambda_{CB})}) - \sigma_n^2(1 - \lambda_d)(\frac{(\lambda_1 f_3 + \lambda_S) f_1}{\alpha(\lambda_1 f_4 + \lambda_{CB})} - \frac{f_2}{\alpha})}{\frac{(1-\lambda_d) f_2 f_1 (\lambda_1 f_3 + \lambda_S)}{\alpha(\lambda_1 f_4 + \lambda_{CB})} + \frac{(\lambda_d - \lambda_u) f_2^2 - (1-\lambda_u) f_2}{\alpha}}\right)^+. \qquad (30)$$

*C. Dual Problem Solution*

The dual problem of OP2 can be solved by using the subgradient method [13]. Therefore, to obtain the solution, we can start with any initial values for the different Lagrangian multipliers and evaluate the optimal powers, then update the Lagrangian multipliers at the next iteration as

$$\lambda_S^{t+1} = \lambda_S^t - \delta(t)\Big[\bar{P}_S - P_S^*\Big], \qquad (31)$$

$$\lambda_{CB}^{t+1} = \lambda_{CB}^t - \delta(t)\Big[\bar{P}_{CB} - P_{CB}^*\Big], \qquad (32)$$

$$\lambda_1^{t+1} = \lambda_1^t - \delta(t)\Big[I_{th} - \Big(f_3 P_S^* + f_4 P_{CB}^*\Big)\Big], \qquad (33)$$

$$\lambda_u^{t+1} = \lambda_u^t - \delta(t)\Big[\frac{1}{2}\log_2\Big(1 + \frac{P_{R,m}^* f_1}{\sigma_n^2}\Big) - \frac{1}{2}\log_2\Big(1 + \frac{P_S^* f_2}{\sigma_n^2}\Big)\Big] \qquad (34)$$

$$\lambda_d^{t+1} = \lambda_d^t - \delta(t)\Big[\frac{1}{2}\log_2\Big(1 + \frac{P_{R,m}^* f_2}{\sigma_n^2}\Big) + \frac{1}{2}\log_2\Big(1 + \frac{P_S^* f_2}{\sigma_n^2}\Big) \\ - \frac{1}{2}\log_2\Big(1 + \frac{P_S^* f_2 + P_{CB}^* f_1}{\sigma_n^2}\Big)\Big], \qquad (35)$$

where $\delta(t)$ is the step size updated according to the nonsummable diminishing step lengths policy [13]. Using the subgradient method, the updated values of the optimal powers and the Lagrangian multipliers are repeated until convergence. The implementation procedures to solve the OP2 is described in Algorithm 1.

## IV. SUBOPTIMAL ALGORITHM

The optimal solution for our non linear OP2 sometimes is difficult to solve due to its high computational complexity. Therefore, in order to solve the problem efficiently, we propose a low complexity suboptimal approach in discrete domain to find suboptimal solution. In the MAC phase, we need to find the optimal power allocation over the terminals (i.e., $P_S$ and $P_{CB}$) in order to maximize the SR of SN without interfering with the PUs.

In this section, we propose a heuristic GA with discrete number of power levels from zero to the peak power budget. In fact, each terminal can transmit its signals using one of the power levels between $0$ and peak power budget, i.e., $\Big(P_S \in \Big\{0, \frac{\bar{P}_S}{N-1}, \frac{2\bar{P}_S}{N-1}, ..., \frac{(N-2)\bar{P}_S}{N-1}, \bar{P}_S\Big\}\Big)$,

**Algorithm 1** Optimal Power Allocation and Relay Selection

- **Input:** $I_{th}, \bar{P}_S, \bar{P}_{CB}, \bar{P}_R, M, f_1, f_2, f_3, f_4, f_5$.
- $\boldsymbol{R_{max}} = \emptyset$.

**for** $m = 1 : M$ **do**

    - $P_{R,m}^* = \min\left(\bar{P}_R, \frac{I_{th}}{f_5}\right)$.

    - Initialize the Lagrangian multipliers $\boldsymbol{\lambda}$, $P_{CB}$, and $\rho = 0$.

    **while** $\rho = 0$ **do**

        - Solve problem (23) to obtain $P_S^*$, $P_S = P_S^*$.

        - Solve problem (24) to obtain $P_{CB}^*$, $P_{CB} = P_{CB}^*$.

        - Update $\boldsymbol{\lambda}$ using subgradient method based on (31) - (33).

        - **Until** Required precision is satisfied or reach maximum iteration.

    **end while**

    - Find $R_{AF}$ using (18)

    - Initialize the Lagrangian multipliers $\boldsymbol{\lambda}$, and $\rho = 1$.

    **while** $\rho = 1$ **do**

        - Solve problem (30) to obtain $P_S^*$, $P_S = P_S^*$.

        - Solve problem (28) to obtain $P_{CB}^*$.

        - Update $\boldsymbol{\lambda}$ using subgradient method based on (31) - (35).

        - **Until** Required precision is satisfied or reach maximum iteration.

    **end while**

    - Find $R_{DF}$ using (25)

    - $R_{max}^{(m)} = \max(R_{AF}, R_{DF})$

**end for**

- Find $m^*$ s.t $R_{opt} = \max\limits_{m} \boldsymbol{R_{max}}$

and $\left(P_{CB} \in \left\{0, \frac{\bar{P}_{CB}}{N-1}, \frac{2\bar{P}_{CB}}{N-1}, ..., \frac{(N-2)\bar{P}_{CB}}{N-1}, \bar{P}_{CB}\right\}\right)$ where $N$ is the number of quantization levels. In this way, the transmitters have more flexibility to allocate their powers in the case where continuous power distribution is not available. The GA tries to find the optimal binary string that maximizes the SR expressed in (9). At the beginning, we represent the discrete quantization values of $P_S$ and $P_{CB}$ as $N$ binary strings each of length $K$, where[3] $K = \lceil \log_2(N) \rceil$. The binary representation set of $P_S$ and $P_{CB}$ are denoted as $S_S$ and $S_{CB}$, respectively. The algorithm concatenates $S_S$ with $S_{CB}$ to produce an initial population set $S_0$ of $N$ elements and each with $2K$ bits, where the first $K$ bits represent the equivalent binary string for $P_S$ and the last $K$ bits represent the equivalent binary string for $P_{CB}$. Initially, the GA computes the SR of all elements in $S_0$ using (9), then maintain the best $\frac{N}{2}$ strings $\in S$ to the next population and from them, generate $\frac{N}{2}$ new strings by applying crossovers technique to form a new population $S$. This procedure is repeated until reaching convergence (i.e., SR remains constant for several successive iterations) or until reaching the maximum generation number $I$. Details of the proposed GA are given in Algorithm 2.

---

[3]$\lceil x \rceil$ denotes the smallest integer not less than $x$.

---

**Algorithm 2** Proposed Genetic Algorithm

- **Input:** $I_{th}, \bar{P}_S, \bar{P}_{CB}, \bar{P}_R, M, f_1, f_2, f_3, f_4, f_5, I$.
- $\boldsymbol{R}_{max} = \emptyset$.

**for** $m = 1 : M$ **do**
- $P_{R,m}^* = \min\left(\bar{P}_R, \frac{I_{th}}{f_5}\right)$.
- $i = 1$, $\boldsymbol{R}_I = \emptyset$, and generate an initial population set $S$.

  **while** ($i \leq I$ or not converge) **do**

    **for** $n = 1 : N$ **do**

      **if** (interference constraint is satisfied) **then**
      - $R^{(n)} =$ Compute the sum rate using (9).

      **else**
      - $R^{(n)} = 0$.

      **end if**

    **end for**

  - $R_I^{(i)} = \max(R)$.
  - Maintain the best $\frac{N}{2}$ strings $\in S$ to the next population and from them, generate $\frac{N}{2}$ new strings by applying crossovers to form a new population $S$.
  - $i = i + 1;$

  **end while**

  - $R_{max}^{(m)} = \max(\boldsymbol{R}_I)$.

**end for**

- Find $m^*$ s.t $R_{opt} = \max\limits_{m} \boldsymbol{R}_{max}$.

---

The formulated OP2 can be, of course, solved via an ES algorithm by investigating all possible combinations of the transmitters power and select the best combinations that satisfied the interference constraint This algorithm requires $M \sum_{i=0}^{2} \binom{2}{i}(N-1)^i = O(MN^2)$ operations [14]. However, our proposed GA requires $MNI$ operations to reach a suboptimal solution.

In the proposed algorithm, our goal is to maximize the SR of the SN without interfering with the PU. The last step in our proposed algorithm is selecting between the AF and DF protocols depends on the higher achieved SR. Hence, our proposed algorithm is able to reach a suboptimal solution with a considerable complexity saving. In addition to that, simulation results in Section V show that by increasing $N$, our proposed GA achieves almost the same performance as the optimal solution.

## V. Selected Simulation Results

In this section, some selected simulation results are performed to show the benefits of our system. We assume a single cell subject to a small scale Rayleigh fading, consisting of one PU and a SN constituted by one CB, one secondary MU, and $M = 4$ relays. The variance $\sigma_n^2$ is assumed to be equal to $10^{-4}$. We also assume that the transmit power constraint of MU, CB, and each RS are equal to $P_{bar}$. The proposed GA is applied under the following settings: the crossover point is chosen randomly between $1$ and $2K$ for each binary string, and we run the GA at most $10$ times.

The advantage of adaptive relaying strategy is depicted in Fig.2. The adaptive strategy can switch between the DF and AF protocols according to the best performance. It is worth mentioning that, in the high SNR regime, adaptive relaying uses the AF protocol. On the other hand, for the low SNR region, adaptive relaying uses the DF protocol. Fig.2(a) plots the SR versus peak power $P_{bar}$, while Fig.2(b) plots the SR versus interference threshold, for different values of $I_{th}$ and MU peak power, respectively. In general, the results suggest the usage of the AF protocol when both $P_{bar}$ and $I_{th}$ are large. This can be justified by noticing that the SR value of the DF protocol becomes as a bottleneck for the first phase in the high SNR regime.

Fig.3 shows a comparison between the performance of the proposed GA with the optimal and ES solutions. We plot the achieved secondary SR versus $P_{bar}$ for different values of $I_{th} = \{20, 5\}$ dBm and different relaying protocols. We can notice that, in the low $P_{bar}$ region, the proposed GA, the optimal solution, and the ES have almost the same sum rate, while in the high $P_{bar}$ region, a gap between these methods is observed. This gap is increasing with higher $P_{bar}$ values. This is justified by the fact that starting from a certain value of $P_{bar}$ the GA can not supply the selected relay with the whole power budget. Hence, the selected relay transmits its signal with one of the quantized power levels. In fact, with high values of $P_{bar}$, the constraint (5) can be affected. For this reason, we introduce the discretization set to get more degrees of freedom by increasing $N$ and as such enhance the SR. For instance, Fig.3(a) and Fig.3(b) plot the secondary SR for $I_{th} = 20$ dBm for DF protocol and AF protocol, respectively. It is shown that the GA achieves almost the same SR reached by the optimal solution. However, when $I_{th}$ is reduced, we notice a degradation of the GA performance at large values of $P_{bar}$ as shown in Fig.3(c) and Fig.3(d).

The same interpretation is applied on Fig.4 in which the achieved secondary SR is plotted versus the interference threshold for both relaying protocols. In this figure, for fixed $P_{bar}$ the performance of the GA is close to the optimal solution for large $I_{th}$. One can see that, a gap between the methods is noticed in the low $I_{th}$ region. This can be justified by the fact that in this region the GA can not reach the maximum power budget due the small value of $I_{th}$. Hence, the GA tries to transmit with one of the quantized power levels. However, It can be shown that when $N \to \infty$, the proposed GA achieves the performance of the optimal solution.

## VI. CONCLUSION

This letter presented an optimal power allocation and a relay selection scheme for two way relaying cognitive radio networks using the dual decomposition method. The idea of this scheme is to maximize the SR of the cognitive network taking into consideration protecting the PUs from the interference caused by the SN. Due to the high computational complexity of the optimal solution, a suboptimal heuristic algorithm is presented. The suboptimal solution based on the GA is able to achieve the same performance of both the ES and optimal solutions with a much less complexity. Furthermore, the performance of the DF and AF schemes, and the impact of the power and interference constraints are illustrated for different interference thresholds and peak power constraints. Finally, the advantage of the adaptive relaying protocol is shown by switching between the DF and AF protocols.

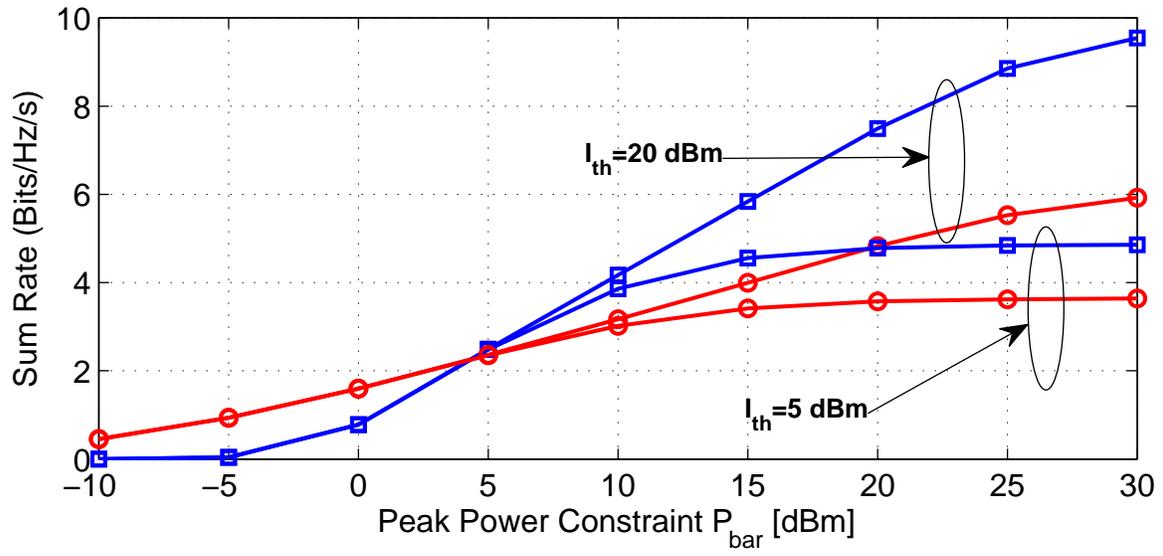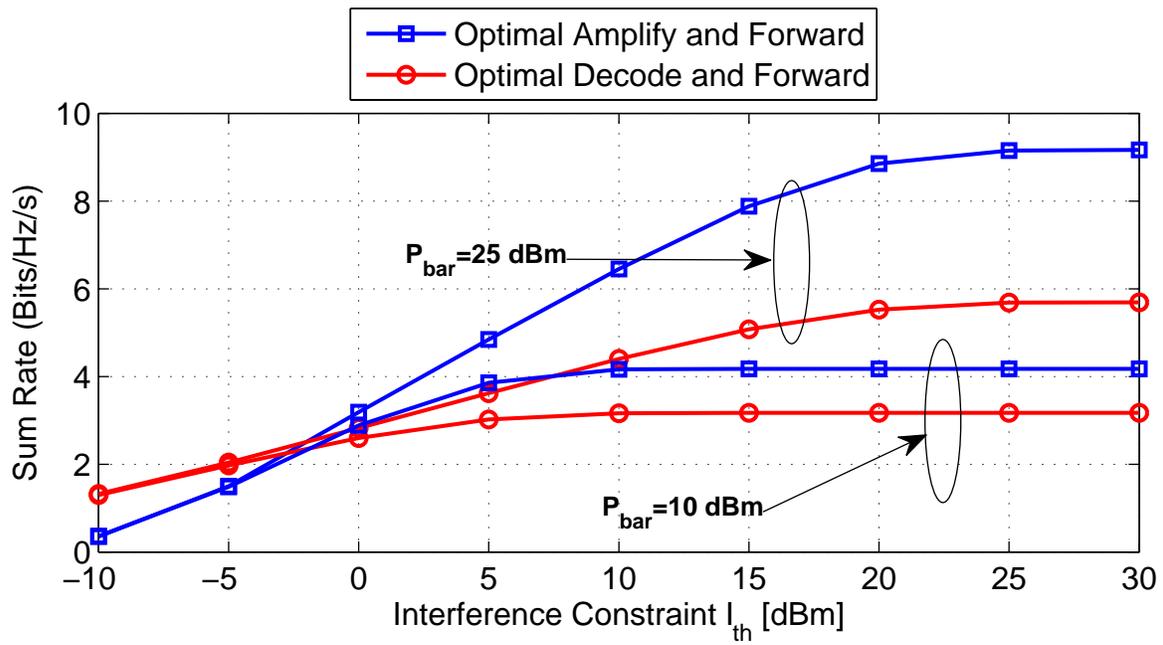

Fig. 2. Achieved SR for the AF ad DF networks versus a) $P_{bar}$, b) $I_{th}$.

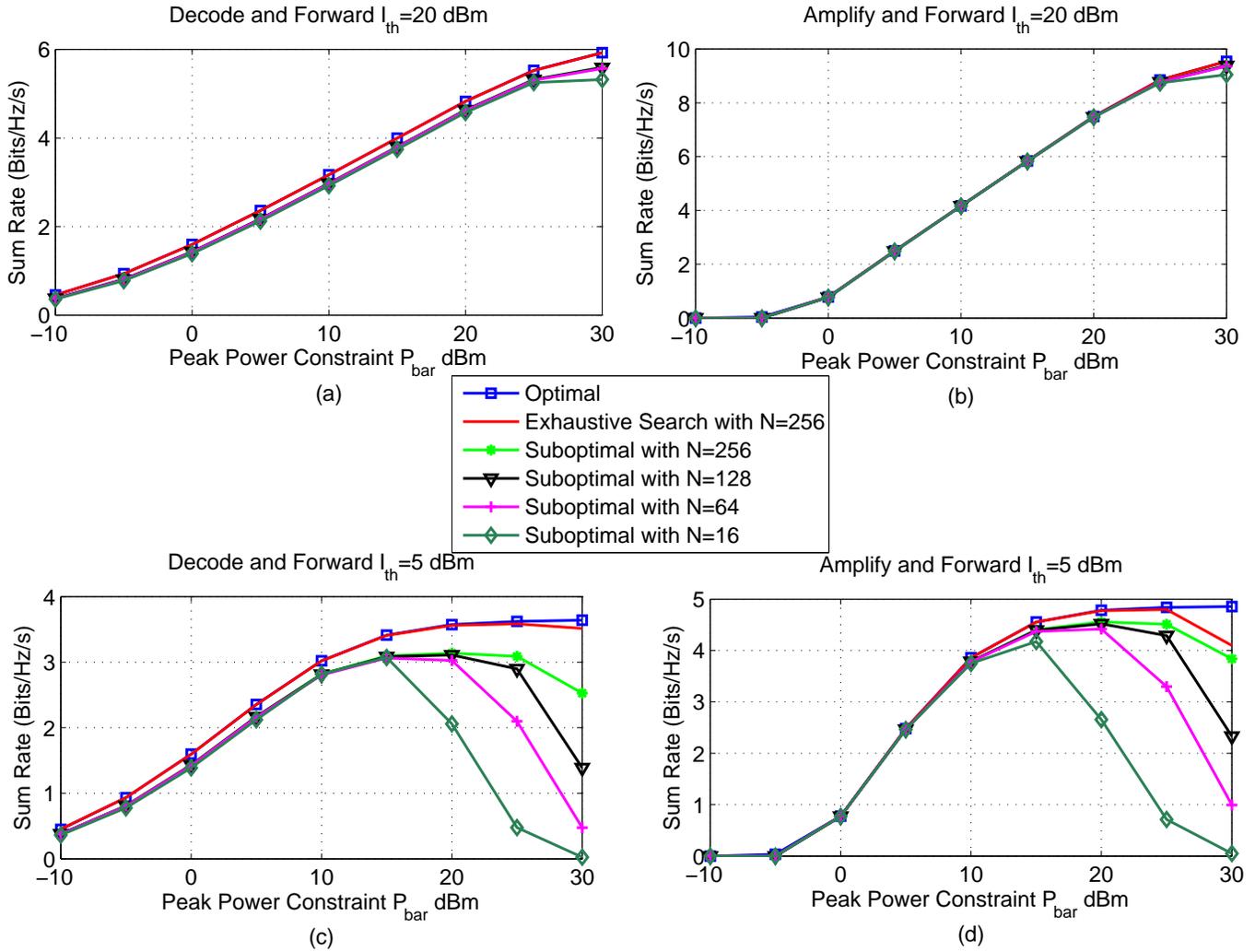

Fig. 3. The achieved SR of the proposed GA, the ES algorithm, and the optimal solution with different values of $I_{th}$, and $N$ versus $P_{bar}$, for (a,c) DF protocol, (b,d) AF protocol.

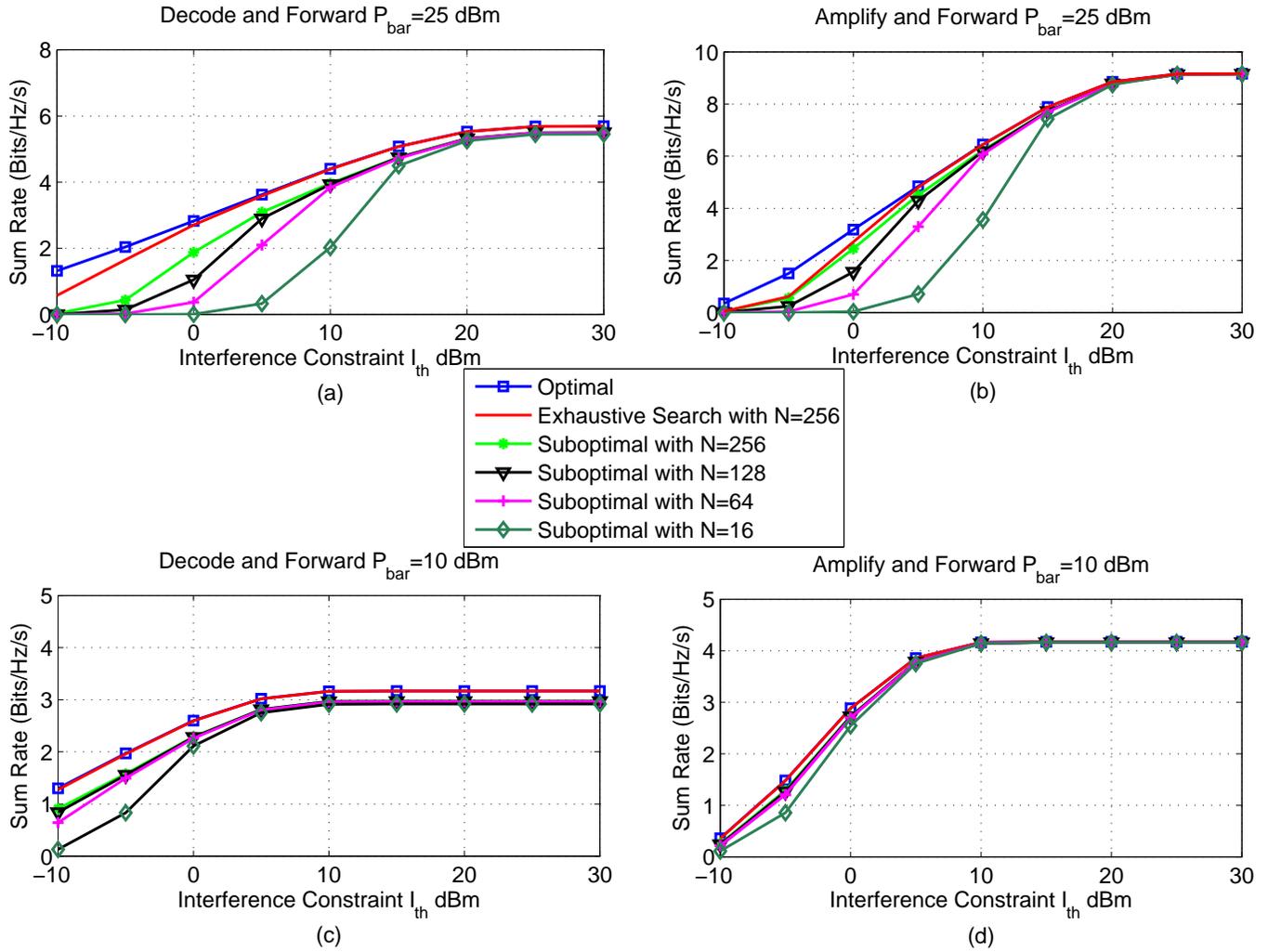

Fig. 4. The achieved SR of the proposed GA, the ES algorithm, and the optimal solution with different values of $P_{bar}$, and $N$ versus $I_{th}$, for (a,c) DF protocol, (b,d) AF protocol.